# Flexocoupling impact on the generalized susceptibility and soft phonon modes in the ordered phase of ferroics

**Anna N. Morozovska[1], Yulian M. Vysochanskii[2], Olexandr V. Varenik[1], Maxim V. Silibin[3], Sergei V. Kalinin[4], and Eugene A. Eliseev[5]**[*]

[1] Institute of Physics, National Academy of Science of Ukraine,
46, pr. Nauky, 03028 Kyiv, Ukraine

[2] Institute of Solid State Physics and Chemistry, Uzhgorod University,
88000 Uzhgorod, Ukraine

[3] National Research University of Electronic Technology "MIET", Bld. 1, Shokin Square,
124498 Moscow, Russia

[4] The Center for Nanophase Materials Sciences, Oak Ridge National Laboratory,
Oak Ridge, TN 37831

[5] Institute for Problems of Materials Science, National Academy of Science of Ukraine,
3, Krjijanovskogo, 03142 Kyiv, Ukraine

## Abstract

Flexoelectric effect impact on the generalized susceptibility and soft phonons dispersion was not studied in the *long-range ordered phases* of ferroics. The gap in the knowledge motivated us to establish the impact of the flexocoupling on the correlation function of the long-range order parameter fluctuations in ferroelectric phase of ferroics with local disordering sources.

Within Landau-Ginzburg-Devonshire approach we obtained analytical expressions for the generalized susceptibility and phonon dispersion relations in the ferroelectric phase. Unexpectedly, the joint action of static and dynamic flexoelectric effect induces non-diagonal components of generalized susceptibility, which amplitude is proportional to the convolution of the spontaneous polarization with flexocoupling constants. The flexocoupling essentially broaden the k-spectrum of generalized susceptibility and so decreases the correlation radii, as well as leads to the additional "pushing away" of the optical and acoustic soft mode phonon branches. The contribution of spontaneous polarization via ferroelectric nonlinearity and electrostriction mechanisms can lead to both broadening and narrowing on the susceptibility k-spectrum. Due to the joint action of flexoelectric coupling and spontaneous polarization the degeneration of the transverse optic and acoustic modes disappears in the soft phonon spectrum in a ferroelectric phase in comparison with the spectra in a paraelectric phase.

These effects have been studied quantitatively for ferroelectric lead zirconate titanate using realistic values of static and dynamic flexocouplings and cubic symmetry approximation for the elastic properties. Also we derived general expressions for correlation function for arbitrary symmetry, elastic and electrostrictive anisotropy. These results can be principally important for quantitative analyses of the experimental data broad spectrum including neutron and Brillouin scattering, which collects unique information from the structural factors and phonon dispersion.

---

[*] Corresponding author: eugene.a.eliseev@gmail.com

# I. Introduction

It is difficult to overestimate the significance of the flexoelectric phenomena contribution to the electromechanics of meso- and especially nanoscale objects, for which the strong strain gradients are inevitable present at the surfaces, interfaces, around point and topological defects [1, 2, 3]. According to experiments and Ginzburg-Landau-type theories, flexoelectricity should strongly influence on the broad spectrum of local electromechanical response of essentially spatially-inhomogeneous systems with inherent strain and/or polarization gradients, ranging from a flexoelectricity-driven imprint [4, 5, 6] and internal bias in thin films [7, 8]; spontaneous flexoelectric effect in nanoferroics [9]; dead layer effect on ferroelectric thin films conditioned by flexoelectricity [10, 11]; the structural, energetic and electro-transport properties of the domain walls and interfaces in ferroelectrics [12, 13, 14, 15, 16] and ferroelastics [17, 18, 19]; hardening of ferroelectrics at nano-indentation [20, 21, 22]; to the local electrochemical strains appeared in response to the excitation of materials with mobile charges by strongly inhomogeneous electric field of the atomic force microscope tip [23, 24] as well as to mechanical writing of ferroelectric polarization by the tip [25]. Notably, flexoelectricity is allowed by symmetry in any material, making the effect widespread and attractive for advanced applications.

Following classical definition the static flexoelectric effect is the response of electric polarization to an elastic strain gradient (direct effect), and, vice versa, the polarization appeared as a response to the strain gradient (inverse effect) [7, 26, 27, 28]. The induced strain is linearly proportional to the polarization gradient $u_{ij}^{sf} = -f_{ijkl}\frac{\partial P_k}{\partial x_l}$, here $f_{ijkl}$ are the components of flexocoupling tensor [1-3], $P_k$ are polarization components. While the static bulk flexoelectric effect can be viewed as an analogue of the piezoelectric effect, the dynamic flexoelectric effect, firstly introduced by Tagantsev as $P_i^{df} = -M_{ij}\frac{\partial^2 U_j}{\partial t^2}$, where $U_j$ is an elastic displacement and $M_{ij}$ is a flexodynamic tensor, has no such analogue, because it corresponds to the polarization response to accelerated motion of the medium in the time domain [1].

Despite the great importance there are only a few ferroics for which the static flexocoupling tensorial coefficients was measured experimentally [29, 30, 31, 32], or obtained from early microscopic estimates [27] or recent *ab initio* calculations [33, 34]. The experimental and theoretical results are rather contradictory, indicating on a limited understanding of the effect nature. The situation with dynamic flexocoupling coefficients is even more unclear. Recently, Kvasov and Tagantsev evaluated the strength of the dynamic flexoelectric effect from *ab initio* calculations and it appeared comparable to that of the static bulk flexoelectric effect [35]. In accordance with this [35] and earlier studies [36, 37] an accurate analysis of the soft phonon spectra extracted from the neutron and Brillouin scattering data can provide information on the components of the total flexocoupling coefficient.

Remarkably that there is an important class of physical properties where the impact of flexocoupling can be critically important is not enough studied, and some aspects are studied rather poorly per se. This is the influence of the static and dynamic flexocouplings on the long-range order parameter fluctuations in the ordered phase of ferroics. Let us underline that the basic experimental methods collecting information about the fluctuations are dynamic dielectric measurements, neutron and Brillouin scattering [38, 39, 40]. Available experimental and theoretical results (see e.g. [41, 42, 43]), mostly demonstrate the significant material-specific impact of the flexocoupling on the scattering spectra. For instance the theory [36-37] predicts a sharp maximum for SrTiO$_3$ in the field dependence of the dielectric loss due to the significant flexoelectric coupling between the soft-mode and acoustic phonon branches, while the analogous field dependence of the loss for Ba$_{0.6}$Sr$_{0.4}$TiO$_3$ appeared monotonic because of small flexoelectric coupling.

Flexoelectric effect impact on the generalized susceptibility and soft phonons dispersion was not studied *theoretically* in the long-range ordered phases of ferroics. The gap in the knowledge motivated us to study the problem for ferroics with local disordering sources (e.g. chemical strains originated from impurity ions or vacancies).

**II. General theory: analytical results near the centre of the Brillouin zone**

Generalized expression for the free energy functional has the following form [24]:

$$F = \int_V d^3r \left( \begin{array}{c} \alpha P_i P_i + \alpha_{ijkl} P_i P_j P_k P_l + \alpha_{ijklmn} P_i P_j P_k P_l P_m P_n + \dfrac{g_{ijkl}}{2}\left(\dfrac{\partial P_i}{\partial x_j}\dfrac{\partial P_k}{\partial x_l}\right) - P_i E_i - q_{ijkl} u_{ij} P_k P_l \\ + \dfrac{c_{ijkl}}{2} u_{ij} u_{kl} + \dfrac{f_{ijkm}}{2}\left(u_{ij}\dfrac{\partial P_m}{\partial x_k} - P_m \dfrac{\partial u_{ij}}{\partial x_k}\right) + \left(\Xi_{ij}(n - n_e) + \beta_{ij}\left(N_d^+ - N_{de}^+\right)\right) u_{ij} \end{array} \right)$$

(1)

Hereinafter summation is performed over all repeating indexes; $P_i$ denotes electric polarization. The expansion coefficient α is temperature dependent, $\alpha = \alpha_T(T - T_C)$, where *T* is the absolute temperature, $T_C$ is the Curie temperature. Elastic strain tensor is $u_{mn}$, $q_{mnij}$ is electrostriction tensor, $f_{mnli}$ is the flexoelectric effect tensor. The higher order coefficients $\alpha_{ijkl}$ and $\alpha_{ijklmn}$ are regarded temperature independent; $g_{ijkl}$ are gradient coefficients tensor, $c_{ijkl}$ are elastic compliances, we introduce the fluctuations $\delta n(\mathbf{r}) = n(\mathbf{r}) - n_e$ and $\delta N_d(\mathbf{r}) = N_d^+(\mathbf{r}) - N_{de}^+$ from the space charge equilibrium values $n_e$ and $N_{de}^+$. Deformation potential tensor is denoted by $\Xi_{ij}$ and Vegard expansion tensor is $\beta_{ij}$.

Dynamic equations of state can be derived from the minimization of Lagrange function, $L = F - T$, where the kinetic energy $T$ is given by expression $T = \frac{\mu}{2}\left(\frac{\partial P_i}{\partial t}\right)^2 + M_{ij}\frac{\partial P_i}{\partial t}\frac{\partial U_j}{\partial t} + \frac{\rho}{2}\left(\frac{\partial U_i}{\partial t}\right)^2$, which includes the dynamic flexoelectric coupling with the tensorial strength $M_{ij}$ [2], $U_i$ is elastic displacement and $\rho$ is the density of a ferroelectric. Corresponding time-dependent Landau-Ginzburg-Devonshire-type equation of state for ferroelectric polarization reads:

$$\mu\frac{\partial^2 P_i}{\partial t^2} + M_{ij}\frac{\partial^2 U_j}{\partial t^2} + 2(\alpha\delta_{ij} - u_{mn}q_{mnij})P_j + 4\alpha_{ijkl}P_jP_kP_l + 6\alpha_{ijklmn}P_jP_kP_lP_mP_n \\ - g_{ijkl}\frac{\partial^2 P_k}{\partial x_j\partial x_l} = f_{mnli}\frac{\partial u_{mn}}{\partial x_l} + E_i \quad (2)$$

The total field is the sum of depolarization (*d*) and small probing external (*ext*) fields, $E_i = E_i^d + \delta E_i^{ext}$. The field should be found self-consistently from the electric potential $\varphi$ as $E_k = -\partial\varphi/\partial x_k$, since the potential satisfy Poisson equation,

$$\varepsilon_b\varepsilon_0\frac{\partial^2\varphi}{\partial x_i^2} = \frac{\partial P_i}{\partial x_i} + e(N_d^+ - n), \quad (3)$$

where $\varepsilon_b$ is background permittivity [44] and $\varepsilon_0$=8.85×10$^{-12}$ F/m is the dielectric permittivity of vacuum, $e(N_d^+ - n)$ the space charge density, $e$=1.6×10$^{-19}$ C the electron charge, $n$ the concentration of the electrons in the conduction band and $N_d^+$ the concentration of ionized donors, e.g. impurity ions or oxygen vacancies.

Elastic strains $u_{ij}$ and stresses $\sigma_{ij}$ are related via Generalized Hook law, which include conventional Hook relation, deformation and chemical stresses, flexoelectric and electrostriction terms [23-24]. Since the time-dependent equation of mechanical equilibrium, $\partial\sigma_{ij}/\partial x_j = \rho\partial^2 U_i/\partial t^2$, should be valid, the equation transforms into dynamic Lame-type equation for elastic strain

$$c_{ijkl}\frac{\partial^2 U_l}{\partial x_j\partial x_k} - \rho\frac{\partial^2 U_i}{\partial t^2} - M_{ij}\frac{\partial^2 P_j}{\partial t^2} = -\frac{\partial}{\partial x_j}\left(\Xi_{ij}\delta n + \beta_{ij}\delta N_d + f_{ijkl}\frac{\partial P_l}{\partial x_k} - q_{ijkl}P_kP_l\right). \quad (4)$$

In order to derive expression for the linear generalized susceptibility and correlation function, let us linearize Eqs.(2) for polarization and Eq. (4) for the displacement in the vicinity of spontaneous values $u_{kl} = u_{kl}^{(s)} + \delta u_{kl}$ and $P_i = P_i^{(s)} + \delta P_i$, where $u_{mn}^{(s)} = s_{mnij}q_{ijkl}P_k^{(s)}P_l^{(s)}$ is the spontaneous strain related to spontaneous polarization $P_l^{(s)}$. Both spontaneous strain and polarization are supposed to be coordinate and time independent in the considered bulk system. Electric field $E_i = E_i^{(s)} + \delta E_i^d + \delta E_i^{ext}$, where depolarization field fluctuations $\delta E_i^d$ will be estimated in Debye approximation as described in **Appendix A** of **Suppl. Mat** [45]**.**

The Fourier representations of the linearized solution for polarization and strain fluctuation in the spatial wave vector **k** and time frequency ω domain have the form:

$$\delta \widetilde{P}_j(\mathbf{k},\omega) = \left(\delta \widetilde{E}_i^{ext} + ik_{j'} S_{mi'}(\mathbf{k},\omega)\left(2ik_n q_{mnij} P_j^{(s)} + f_{mnli} k_l k_n - M_{mi}\omega^2\right)\delta \widetilde{C}_{i'j'}\right)\widetilde{\chi}_{ij}(\mathbf{k},\omega), \quad (5a)$$

$$\delta \widetilde{U}_k(\mathbf{k},\omega) = -ik_j \delta \widetilde{C}_{ij} S_{ik}(\mathbf{k},\omega) + S_{ik}(\mathbf{k},\omega)\widetilde{\chi}_{sl}(\mathbf{k},\omega)\left(f_{ijml} k_j k_m - M_{il}\omega^2 - 2ik_j q_{ijnl} P_n^{(s)}\right) \\ \times \left(\delta \widetilde{E}_s^{ext} + i\left(f_{qnps} k_p k_n - M_{qs}\omega^2\right) k_{j'} S_{qi'}(\mathbf{k},\omega)\delta \widetilde{C}_{ij}\right) \quad (5b)$$

Where $\delta \widetilde{C}_{ij} = \left(\Xi_{ij}\delta \widetilde{n} + \beta_{ij}\delta \widetilde{N}_d\right)$. Since the harmonic approach (5) is applicable for small **k,** we would like to underline that we did not aim to reach the quantitative agreement between the calculated and experimentally observed soft phonon spectra entire the first Brillouin zone. Consideration of the problem for higher **k** values requires including of the anharmonicity and higher gradient terms [46].

The inverse matrices of the Green tensor, $\widetilde{\chi}_{ij}(\mathbf{k},\omega)$, that is in fact correlation function or generalized susceptibility $\left.\dfrac{\partial \widetilde{P}_i(\mathbf{k},\omega)}{\partial E_j^{ext}}\right|_{E_j^{ext}\to 0} \equiv \widetilde{\chi}_{ij}(\mathbf{k},\omega)$, and elastic function $S_{ir}(\mathbf{k},\omega)$ are

$$\widetilde{\chi}_{ij}^{-1}(\mathbf{k},\omega) = \beta_{ij}(\mathbf{k},\omega) + \Theta_{ipjl}(\mathbf{k},\omega) + Q_{ij}(\mathbf{k},\omega) + \gamma_{ijkl}(\mathbf{k},\omega)P_k^{(s)}P_l^{(s)}, \quad (6a)$$

$$S_{ik}^{-1}(\mathbf{k},\omega) = c_{ijkl} k_l k_j - \rho\omega^2 \delta_{ik}. \quad (6b)$$

Here the linear dynamic stiffness is affected by depolarization effect as $\beta_{ij}(\mathbf{k},\omega) = (2\alpha - \mu\omega^2)\delta_{ij} + \dfrac{k_i k_j}{\varepsilon_b \varepsilon_0 (k^2 + R_d^{-2})}$, where $R_d$ is a Debye screening radius. Nonlinear stiffness $\gamma_{ijkl}(\mathbf{k},\omega) = 12\alpha_{ijkl} - 2q_{mnij}q_{i'j'kl}s_{mni'j'} - 4q_{mnil}q_{i'j'jk}k_j k_n S_{mi'}(\mathbf{k},\omega) + 30\alpha_{ijklmn}P_m^{(s)}P_n^{(s)}$. The flexoelectric coupling changes the polarization gradient coefficient tensor $g_{ipjl}$ to the frequency ω and k-dependent tensorial function that was introduced as $\Theta_{ipjl}(\mathbf{k},\omega) = g_{ipjl} k_p k_l - (f_{mnli} k_n k_l - M_{mi}\omega^2)(f_{i'j'pj} k_j k_p - M_{i'j}\omega^2)S_{mi'}(\mathbf{k},\omega)$. A new complex term $Q_{ij}(\mathbf{k},\omega)$ is proportional to the convolution of the spontaneous polarization vector with the static and dynamic flexocoupling constants:

$$Q_{ij}(\mathbf{k},\omega) = 2iS_{mi'}(\mathbf{k},\omega)\left((f_{mnpi} k_p k_n - M_{mi}\omega^2)k_{j'} q_{i'j'kj} - (f_{i'j'pj} k_p k_{j'} - M_{i'j}\omega^2)k_n q_{mnik}\right)P_k^{(s)}. \quad (7)$$

The term is absent in the paraelectric phase. Here the static and dynamic flexocoupling appeared in a universal combination $\left(f_{mnpi}k_p k_n - M_{mi}\omega^2\right)$.

Order parameter correlation function is related with the generalized susceptibility via Callen-Welton [47] fluctuation-dissipation theorem and corresponding correlations radius can be determined from direct matrix $\widetilde{\chi}_{ij}(\mathbf{k},\omega)$. In general case analytical expressions for $\widetilde{\chi}_{ij}(\mathbf{k},\omega)$ are rather cumbersome. In order to analyze analytically a concrete case, below we consider a uniaxial ferroelectric with a spontaneous polarization directed along z-axes, $\mathbf{P}^{(s)} = (0,0,P_S)$ and other tensorial properties (elastic, electrostrictive and flexoelectric) in the *cubic symmetry approximation*.

Following Cochren papers [38], dynamical structural factor of neutron scattering is proportional to the dynamic susceptibility spectra $\widetilde{\chi}(\mathbf{k},\omega)$. Integral intensity of the scattering is proportional to the static spectra, $(d\sigma/d\Omega)\sim \widetilde{\chi}(\mathbf{k},0)$. In the next section we discuss the influence of the flexocoupling on the static spectrum of dielectric susceptibility in a ferroelectric phase.

**III. Flexocoupling impact on the dynamic generalized susceptibility in a ferroelectric phase**

For the **Case I**, when a wave vector $\mathbf{k}=(0,0,k_z)$ is longitudinal with respect to the spontaneous polarization direction $\mathbf{P}_S = (0,0,P_S)$, corresponding nonzero components of the susceptibility are [45]:

$$\widetilde{\chi}_{11}(\mathbf{k},\omega) = \widetilde{\chi}_{22}(\mathbf{k},\omega) = \frac{1}{\alpha_{11}^* - \mu\omega^2 + g_{44}^* k_z^2}. \tag{8a}$$

$$\widetilde{\chi}_{33}(\mathbf{k},\omega) = \left(\alpha_{33}^* - \mu\omega^2 + g_{11}^* k_z^2 + \frac{k_z^2}{\varepsilon_b \varepsilon_0 \left(k_z^2 + R_d^{-2}\right)}\right)^{-1}. \tag{8b}$$

The spontaneous polarization contributes to the spectra of $\widetilde{\chi}_{ij}(\mathbf{k},\omega)$ via the renormalization of the dielectric stiffness coefficient $\alpha$ as $\alpha_{11}^*(\mathbf{k},\omega) = 2\alpha + P_S^2\left(\beta_{12}^* - \frac{q_{44}^2 k_z^2}{c_{44}k^2 - \rho\omega^2}\right) + 2\alpha_{112}P_S^4$ and $\alpha_{33}^*(\mathbf{k},\omega) = 2\alpha + P_S^2\left(\beta_{11}^* - \frac{4q_{11}^2 k_z^2}{c_{11}k_z^2 - \rho\omega^2}\right) + 30\alpha_{111}P_S^4$. Nonlinear stiffness $\alpha_{ijkl}$ is renormalized by electrostriction coupling as $\beta_{11}^* = 12\alpha_{11} - \frac{2(q_{11}+2q_{12})^2}{3(c_{11}+2c_{12})} - \frac{2q_{44}^2}{3c_{44}}$ and $\beta_{12}^* = 2\alpha_{12} - \frac{2(q_{11}+2q_{12})^2}{3(c_{11}+2c_{12})} + \frac{q_{44}^2}{3c_{44}}$. Thus the contribution of spontaneous polarization via ferroelectric nonlinearity ($\sim \beta_{12}^* P_S^2$) and electrostriction ($\sim q_{ij} q_{kj} P_S^2$) mechanisms can lead to either increase or decrease of the coefficients $\alpha_{ij}^*$ depending on the material constants signs.

Flexocoupling changes the gradient coefficients as

$$g_{11}^*(\mathbf{k},\omega) = g_{11} - \frac{\left(f_{11} k_z^2 - M_{11} \omega^2\right)^2}{k_z^2 \left(c_{11} k_z^2 - \rho \omega^2\right)}, \quad g_{44}^*(\mathbf{k},\omega) = g_{44} - \frac{\left(f_{44} k_z^2 - M_{11} \omega^2\right)^2}{k_z^2 \left(c_{44} k_z^2 - \rho \omega^2\right)}. \quad (8c)$$

The term $(\varepsilon_b \varepsilon_0)^{-1} k_z^2 / (k_z^2 + R_d^{-2})$ in $\tilde{\chi}_{33}$ originates from the depolarization electric field.

The static k-spectra of $\tilde{\chi}_{ij}(\mathbf{k},0)$ calculated with and without flexocoupling contribution are shown in the **Figure 1a**. The component $\tilde{\chi}_{33}$ is much smaller that the ones due to the depolarization effect. As one can see from the figure flexoelectric effect essentially broadens k-spectrum of all susceptibility components and the broadening increases with k increase. Both spectra coincide in the point **k**=0 as anticipated. The dynamic flexoeffect does not contribute to the spectra in the static case ($\omega$=0). The contribution of spontaneous polarization via ferroelectric nonlinearity and electrostriction mechanisms can lead to both broadening and narrowing on the different components of susceptibility k-spectra.

For the **Case II**, when a fluctuation wave vector $\mathbf{k} = (k_x, 0, 0)$ is transverse with respect to the spontaneous polarization direction $\mathbf{P}_S = (0, 0, P_S)$, corresponding nonzero components of the generalized susceptibility in Voight notations are [45]:

$$\tilde{\chi}_{11}(\mathbf{k},\omega) = \frac{\alpha_{11}^* - \mu \omega^2 + g_{44}^* k_x^2}{\Delta_{22}(\mathbf{k},\omega)}, \quad \tilde{\chi}_{22}(\mathbf{k},\omega) = \frac{1}{\alpha_{22}^* - \mu \omega^2 + g_{44}^* k_x^2}, \quad (9a)$$

$$\tilde{\chi}_{33}(\mathbf{k},\omega) = \left(\alpha_{33}^* - \mu \omega^2 + g_{11}^* k_x^2 + \frac{k_x^2}{\varepsilon_b \varepsilon_0 \left(k_x^2 + R_d^{-2}\right)}\right) \frac{1}{\Delta_{22}(\mathbf{k},\omega)}, \quad (9b)$$

$$\tilde{\chi}_{13}(\mathbf{k},\omega) = -\tilde{\chi}_{31}(\mathbf{k},\omega) = \frac{-2i P_S k_x}{\Delta_{22}(\mathbf{k},\omega)} \left(\frac{q_{12} \left(f_{11} k_x^2 - M_{11} \omega^2\right)}{c_{11} k_x^2 - \rho \omega^2} - \frac{q_{44} \left(f_{44} k_x^2 - M_{11} \omega^2\right)}{2 \left(c_{44} k_x^2 - \rho \omega^2\right)}\right). \quad (9c)$$

The spontaneous polarization contributes to the components by the renormalization of the linear dielectric stiffness coefficients $\alpha_{11}^*(\mathbf{k},\omega) = 2\alpha + \left(\beta_{11}^* - \frac{4 q_{12}^2 k_x^2}{c_{11} k_x^2 - \rho \omega^2}\right) P_S^2 + 30 \alpha_{111} P_S^4$,

$\alpha_{22}^*(\mathbf{k},\omega) = 2\alpha + \beta_{12}^* P_S^2 + 2\alpha_{112} P_S^4$ and $\alpha_{33}^*(\mathbf{k},\omega) = 2\alpha + \left(\beta_{12}^* - \frac{q_{44}^2 k_x^2}{c_{44} k_x^2 - \rho \omega^2}\right) P_S^2 + 2\alpha_{112} P_S^4$. The form of the gradient functions $g_{11}^*(\mathbf{k},\omega)$ and $g_{44}^*(\mathbf{k},\omega)$ used in Eqs.(9a)-(9b) are the same as in the longitudinal case with the only substitution $k_z \to k_x$ in Eq.(8c). Note that nonzero non-diagonal element $\tilde{\chi}_{13}(\mathbf{k},\omega)$ is proportional to the product of the spontaneous polarization value and flexocoupling constants. Denominator $\Delta_{22}(\mathbf{k},\omega)$ is expressed in terms of inverse matrix elements, $\Delta_{22}(\mathbf{k},\omega)=\tilde{\chi}_{11}^{-1}(\mathbf{k},\omega)\tilde{\chi}_{33}^{-1}(\mathbf{k},\omega)-\tilde{\chi}_{13}^{-1}(\mathbf{k},\omega)\tilde{\chi}_{31}^{-1}(\mathbf{k},\omega)$. The evident expression for $\Delta_{22}(\mathbf{k},\omega)$ is:

$$\Delta_{22}(\mathbf{k},\omega) = -4k_x^2 P_S^2 \left( \frac{q_{12}\left(f_{11}k_x^2 - M_{11}\omega^2\right)}{c_{11}k_x^2 - \rho\omega^2} - \frac{q_{44}\left(f_{44}k_x^2 - M_{11}\omega^2\right)}{2\left(c_{44}k_x^2 - \rho\omega^2\right)} \right)^2 + \\ \left(\alpha_{11}^* - \mu\omega^2 + g_{44}^* k_x^2\right)\left(\alpha_{33}^* - \mu\omega^2 + g_{11}^* k_x^2 + \frac{k_x^2}{\varepsilon_b \varepsilon_0 \left(k_x^2 + R_d^{-2}\right)}\right)$$

(9d)

The flexocoupling induces several principal changes in the susceptibilities, in particular the terms directly proportional to the product of spontaneous polarization and flexocoupling constants originated from $\tilde{\chi}_{13}^{-1}(\mathbf{k},\omega)$, as well as to the changes related with the gradient functions $g_{ii}^*(\mathbf{k},\omega)$.

The static k-spectra of $\tilde{\chi}_{ij}(\mathbf{k},0)$ calculated with and without flexocoupling contribution for ferroelectric PZT are shown in the **Figure 1b**. The strong inequalities $\tilde{\chi}_{11} \ll \tilde{\chi}_{33} \ll \tilde{\chi}_{22}$ and $|\tilde{\chi}_{13}| \ll \tilde{\chi}_{33}$ valid due to the depolarization effect, because the denominator $\Delta_{22}(\mathbf{k},0)$ includes the depolarization factor $(\varepsilon_b\varepsilon_0)^{-1}k^2/(k^2+R_d^{-2})$ and thus strongly decreases $\tilde{\chi}_{11}$, $\tilde{\chi}_{13}$ and $\tilde{\chi}_{33}$ in comparison with the component $\tilde{\chi}_{22}$, which is not affected by depolarizing effect at all as it should be for transverse fluctuations of polarization z-component in the direction 1. Since $\tilde{\chi}_{33}$ contains the depolarization factor in the numerator it becomes much higher than the components $\tilde{\chi}_{11}$ and $\tilde{\chi}_{13}$. As one can see from the figure flexoelectric effect induces the non-diagonal component $\tilde{\chi}_{13}$, that is odd with respect to **k**, and essentially broadens k-spectrum of the susceptibility diagonal components. The broadening increases with k increase.

Correlation radii tensor $R_{ij}$ is proportional to the second derivative of the generalized susceptibility, $R_{ij}^2 = -\frac{1}{2\tilde{\chi}_{ij}}\left(\frac{\partial^2 \tilde{\chi}_{ij}(\mathbf{k},0)}{\partial k^2}\right)^2\bigg|_{k\to 0}$, where k is either $k_z$ or $k_x$. The dependences of correlation radii of $R_{ij}$ on flexoelectric coefficients $f_{11}$ and $f_{44}$ are shown in the **Figure 1c** and **1d** correspondingly. The correlation radii either monotonically decrease with the flexoelectric coupling constants $f_{ij}$ increase or remained independent on the some $f_{ij}$. In particular $R_{13}$ always decreases $f_{ij}$ increase, because $\tilde{\chi}_{13}$ is proportional to $f_{ij}$. The situation with other $R_{ij}$ depends on the orientation of **k** vector with respect to the polarization **P**.

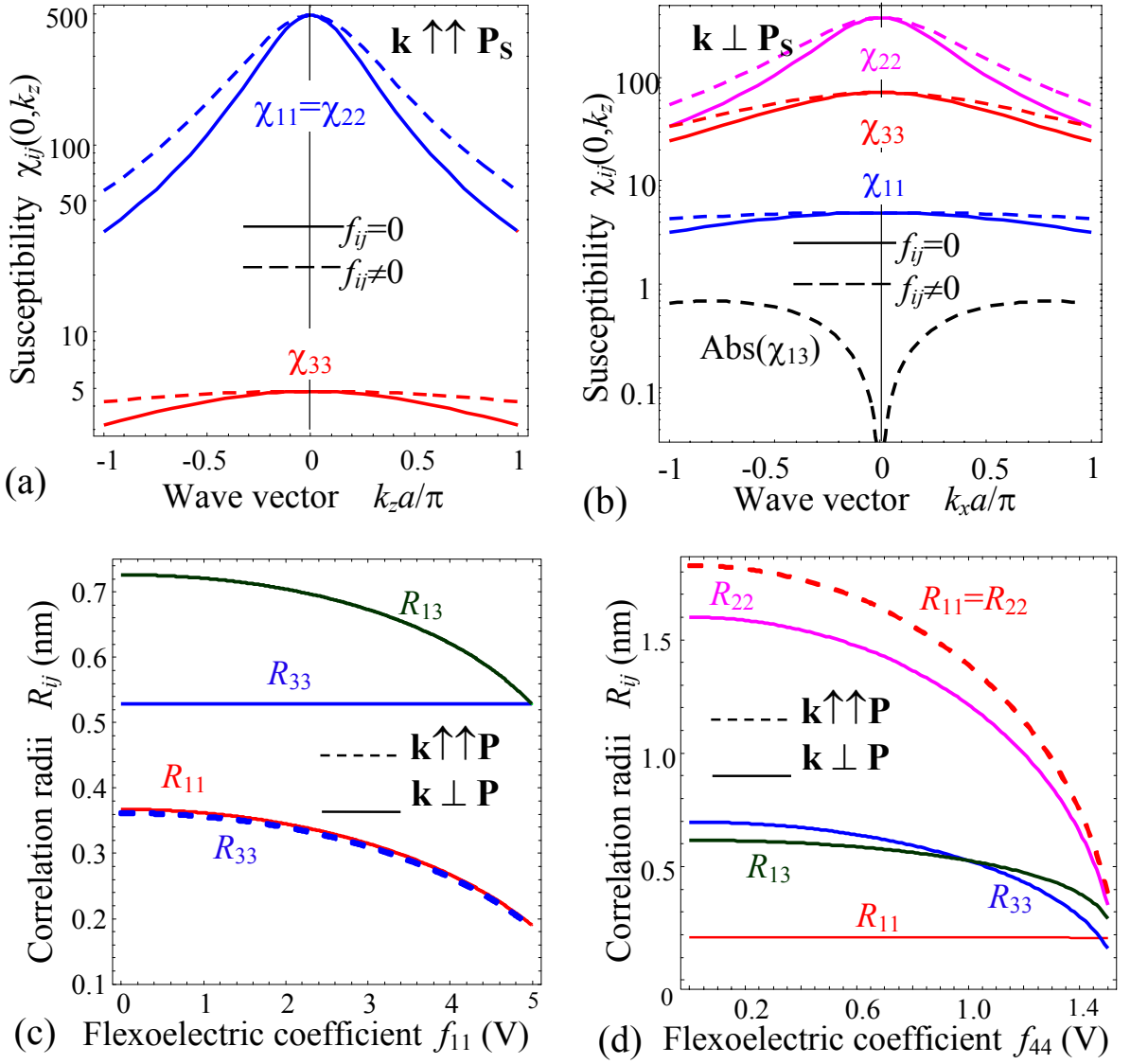

**Figure 1**. Spatial spectrum of the generalized susceptibility nonzero components $\tilde{\chi}_{ij}(\omega=0,k_x)$ vs. $k$ in units $\pi/a$ calculated for longitudinal $\vec{k} \uparrow\uparrow \vec{P}$ **(a)** and transverse $\vec{k} \perp \vec{P}$ **(b)** fluctuation wave vector directions with respect to the spontaneous polarization $\mathbf{P}=(0,0,P_S)$, $a$ is a lattice constant. Dashed curves are calculated with flexoelectric effect and solid curves without it. **(c,d)** Correlation radii $R_{ij}$ dependence on the flexoelectric coefficient $f_{11}$ **(c)** and $f_{44}$ **(d)**. Solid curves are calculated for $\vec{k} \perp \vec{P}$ and dashed curves correspond to $\vec{k} \uparrow\uparrow \vec{P}$ wave vector directions. Parameters corresponding to PZT are listed in the **Table 1,** room temperature $T$=300 K**.**

Table I. Material parameters for bulk ferroelectric

| coefficient | PbZr$_{0.4}$Ti$_{0.6}$O$_3$ (from [48, 49,]) | PbTiO$_3$ (from [50]) |
|---|---|---|
| $\varepsilon_b$ | 5  [44] | 5 |
| $\alpha_{iT}$ ($\times 10^5$C$^{-2}\cdot$mJ/K) | 2.12 | 3.765 |
| $T_C$  (K) | 691 | 752 |
| $\alpha^{(\sigma)}_{ij}$ ($\times 10^8$C$^{-4}\cdot$m$^5$J) | $a_{11}$= 0.3614, $a_{12}$= 3.233 | $a_{11}$= − 0.725, $a_{12}$=7.50 |
| $\alpha_{ijk}$ ($\times 10^8$ Jm$^9$C$^{-6}$·) | $a_{111}$= 1.859, $a_{112}$= 8.503 $a_{123}$= −40.63 | $a_{111}$= 2.606, $a_{112}$= 6.10, $a_{123}$= −36.60 |
| $q_{ij}$ ($\times 10^9$ V·m/C) | $q_{11}$=8.91, $q_{12}$= −0.787, $q_{44}$=3.18 | $q_{11}$=11, $q_{44}$=7 |
| $c_{ij}$ ($\times 10^{10}$ Pa) | $c_{11}$=17.0, $c_{12}$=8.2, $c_{44}$=4.7 | $c_{11}$=18, $c_{12}$=7.9, $c_{44}$=11 |
| $g_{ij}$ ($\times 10^{-10}$C$^{-2}$m$^3$J) | $g_{11}$=2.0, $g_{44}$=1.0 * Estimated form domain wall width | $g_{11}$=1.5, $g_{44}$=0.5 |
| $f_{ij}$ (V) | $f_{11}$= 5, $f_{12}$= −1, $f_{44}$=+1 *estimated from [27, 30, 31, 32] | $f_{11}$= −8, $f_{44}$= −1.9 |
| $M_{11}$ (V s$^2$/m$^2$) | $6\times10^{-8}$    [35] | $-2\times10^{-8}$ |
| $\rho$ ($\times 10^3$ kg/m$^3$) | 8.087 * *At normal conditions | 7.986 |
| $\mu$ ($\times 10^{-18}$ s$^2$mJ) | 1.413    [41] | 1.59 |
| $R_d$ (m) | from 20 nm to infinity | infinity |

### IV. The impact of flexocoupling on soft phonon spectra in ferroelectric phase

Starting from early Shirane papers [41-43] soft phonon branches were studied experimentally and fitted theoretically in a high-temperature paraelectric phase for several incipient and proper ferroelectrics. Below we study the impact of the flexocoupling on soft phonon spectra in ferroelectric phase and compare the results with a paraelectric phase.

Dispersion relations for longitudinal and transverse optical (LO and TO) and acoustical (LA and TA) modes can be obtained from the analyses of the determinant $\det[\tilde{\chi}_{ij}^{-1}(\mathbf{k},\omega)]=0$. Dispersion relation for the fluctuation wave vector direction $\mathbf{k}=(0,0,k_z)$ for the cases $\mathbf{k}\uparrow\uparrow\delta P$ and $\mathbf{k}\perp\delta P$ are

$$\begin{pmatrix} 2\alpha-\mu\omega^2+g_{11}k_z^2-\dfrac{(f_{11}k_z^2-M_{11}\omega^2)^2}{c_{11}k_z^2-\rho\omega^2}+\dfrac{k_z^2}{\varepsilon_b\varepsilon_0(k_z^2+R_d^{-2})} \\ +P_S^2\left(\beta_{11}^*-\dfrac{4q_{11}^2k_z^2}{c_{11}k_z^2-\rho\omega^2}\right)+30\alpha_{111}P_S^4 \end{pmatrix}=0, \quad (10a)$$

$$2\alpha-\mu\omega^2+g_{44}k_z^2-\dfrac{(f_{44}k_z^2-M_{11}\omega^2)^2}{c_{44}k_z^2-\rho\omega^2}+P_S^2\left(\beta_{12}^*-\dfrac{q_{44}^2k_z^2}{c_{44}k_z^2-\rho\omega^2}\right)+2\alpha_{112}P_S^4=0 \quad (10b)$$

Dispersion relation for the fluctuation wave vector direction $\mathbf{k}=(k_x,0,0)$ has the form

$$\left(2\alpha-\mu\omega^2+g_{44}^*(\mathbf{k},\omega)k_x^2+P_S^2\beta_{12}^*\right)\Delta_{22}(\mathbf{k},\omega)=0. \quad (10c)$$

The terms originated from the static and dynamic flexocoupling appeared in the combination $(f_{11}k_z^2-M_{11}\omega^2)$ and $(f_{44}k_z^2-M_{11}\omega^2)$ in the equations. The spontaneous polarization $P_S$ via ferroelectric nonlinearity and electrostriction mechanisms generate the term proportional to $\beta_{ij}^*P_S^2$, $\alpha_{ijkl}P_S^4$ and $q_{ij}q_{kj}P_S^2$ in the equations. Due to the k-dependence of the terms $\sim P_S^2$ the analytical solution of Eqs.(10) is absent in a ferroelectric phase.

The features of the soft phonon k-spectra were calculated with static ($f_{ij} \neq 0$ and $M_{ij}=0$) and dynamic flexocoupling ($f_{ij} \neq 0$ and $M_{ij} \neq 0$) and without it ($f_{ij}=0$ and $M_{ij}=0$). Spectra calculated in the paraelectric and ferroelectric phase for the cases $\vec{k} \uparrow\uparrow \vec{P}_S$ and $\vec{k} \perp \vec{P}_S$ are compared in the **Figure 2a, 2b** and **2c** correspondingly. Parameters corresponding to PZT are listed in the **Table I.**

Equations (10) have relatively simple analytical solution in a paraelectric phase ($P_S^2 = 0$), namely two acoustic (LA and TA) and two optical (LO and TO) modes (see **Figure 2a**). Equation (10a) has an analytical solution in a ferroelectric phase also, and it contains a very high longitudinal optical mode (LO) with frequency at about $150 \times 10^{12} \text{s}^{-1}$ in the dielectric limit ($R_d^{-2} \to 0$) and one acoustic mode LA$_3$. The LO-mode is weakly dependent on temperature due to the depolarization factor $\dfrac{k_z^2}{\varepsilon_b \varepsilon_0 (k_z^2 + R_d^{-2})}$, that becomes giant in the considered dielectric limit. Both paraelectric and ferroelectric spectra contain rather high frequency longitudinal optic modes (LO) due to the strong depolarization field, that is maximal in the dielectric limit ($R_d^{-2} \to 0$) and is almost independent on the flexocouplings and temperature. Therefore the LO modes are not shown in the **Figure 2.** The longitudinal soft mode is insensitive to the flexocoupling, because its dispersion is strongly affected by the depolarization effect.

Due to the k-dependence of the terms $\sim P_S^2$ the analytical solution of Eqs.(10b) is absent in a ferroelectric phase. Corresponding numerical solution has four degenerated transverse soft phonon branches, two optical (TO$_1$ = TO$_2$) and two acoustic (TA$_1$ = TA$_2$) modes (see **Figure 2b**). All the transverse soft modes are relatively sensitive to both the dynamic and static flexocoupling constants especially at $k_z a/\pi \geq 0.03$, $a$ is a lattice constant (compare solid, dotted and dashed curves for TO modes in **Figure 2a** and **2b**). Since the calculated phonon spectrum in the paraelectric phase has two acoustic (LA and TA) and two optical (LO and TO) modes, we can conclude that the appearance of spontaneous polarization does not lead to the qualitative changes in the spectra for the case of wave vector direction $\vec{k} \uparrow\uparrow \vec{P}_S$.

Without flexocoupling the numerical solution of Eq.(10c) has six different phonon branches in the ferroelectric phase for the case $\vec{k} \perp \vec{P}_S$, namely three optical (LO, TO$_2$ and TO$_3$) and three acoustic modes (LA$_1$, TA$_2$, TA$_3$), at that the frequencies of the modes TA$_2$ and TA$_3$ are almost the same at $ka_x/\pi < 0.3$ (see solid curves in the **Figure 2c**). With the flexocoupling included, the solution in the ferroelectric phase has also six different soft phonon branches, three optical (LO, TO$_2$ and TO$_3$) and three acoustic (LA$_1$, TA$_2$ and TA$_3$) modes, at that the frequencies of the modes TA$_2$ and TA$_3$ are noticeably different at $ka_x/\pi < 0.3$ (see dashed and dotted curves in the **Figure 2c**). Since the phonon spectra in the paraelectric phase has two optical (LO and TO) and two acoustic (LA and TA) modes (see **Figure 2a**), we can conclude that the spontaneous polarization appearance leads to the removal of the degeneration of the acoustic and optic modes TA and TO for the case $\vec{k} \perp \vec{P}_S$ and consequently to the appearance of different transverse acoustic and optics modes TA$_2$ and TA$_3$, TO$_2$ and TO$_3$. The transverse TO$_{2,3}$ and TA$_{2,3}$ modes are relatively sensitive to both static and dynamic flexoelectric coupling strength for the case $\vec{k} \perp \vec{P}_S$, acoustic at $k_x a/\pi \geq 0.1$ and optic all small k, meanwhile the longitudinal LA$_1$ mode becomes sensitive to the coupling at $k_x a/\pi \geq 0.15$ (compare solid, dotted and dashed curves in **the Figure 2c**). The flexoelectric coupling significantly increases the splitting of the TA$_2$ and TA$_3$ modes. Moreover, TO$_3$ and LA$_1$ modes are "pushed away" by the static and dynamic flexocoupling in the ferroelectric phase at small **k** ($k_x a/\pi \leq 0.15$) and start to approach each other at $k_x a/\pi \geq 0.15$ (compare solid and dashed curves inside in the **Figure 2d**). The effects give us the opportunity to define the static and dynamic flexocoupling constants (e.g. $f_{11}$, $f_{44}$ and $M_{11}$) from soft phonons spectra in the assumption of other known materials parameters.

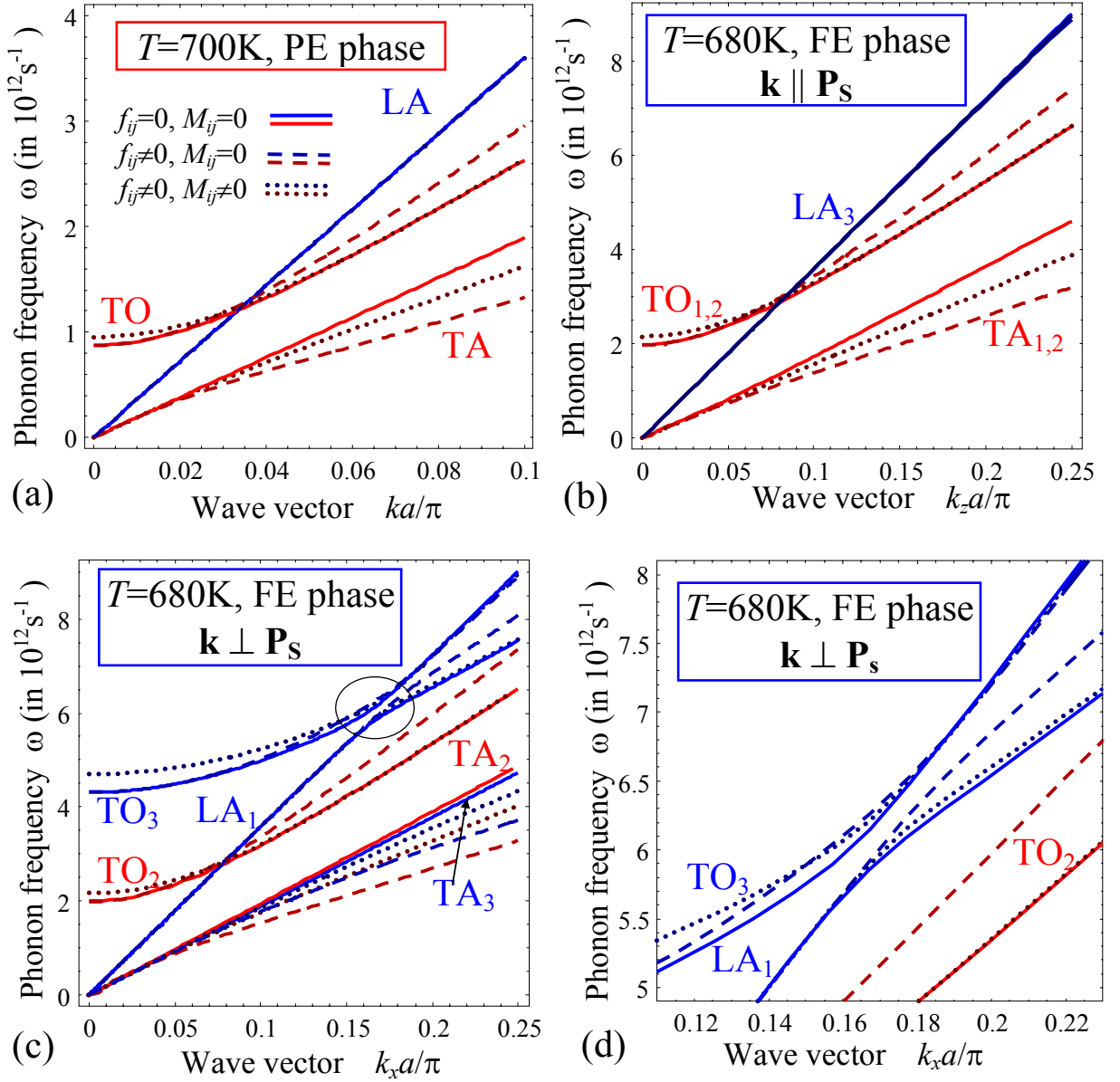

**Figure 2**. Soft phonon branches calculated in PZT vs. $k$ in units $\pi/a$, $a$ is a lattice constant. Plot **(a)** corresponds to the paraelectric phase of PZT (T=700 K), plots **(b)** and **(c)** are calculated in the ferroelectric phase (T=680 K) for longitudinal $\vec{k} \uparrow\uparrow \vec{P}$ **(b)** and transverse $\vec{k} \perp \vec{P}$ **(c,d)** fluctuation of the wave vector directions with respect to the spontaneous polarization $\mathbf{P} = (0, 0, P_S)$. Solid curves are calculated without flexoelectric coupling ($f_{ij}=0$ and $M_{ij}=0$); dashed curves are calculated with the static but without the dynamic effect ($f_{ij}\neq0$ and $M_{ij}=0$) and dotted curves are calculated with the dynamic and static flexoelectric effects included ($f_{ij}\neq0$ and $M_{ij}\neq0$). $TA_2$ and $TA_3$ modes coincide for the zero flexocoupling in the plot **(c)**. **(d)** Zoom of the plot (c) inside the circle. The highest longitudinal optical mode (LO) at about $150 \times 10^{12} s^{-1}$ is not shown in the plots. Parameters corresponding to PZT are listed in the **Table I**.

Finally, let us answer on the question how important is the flexocoupling for quantitative description of the observed phonon spectra. In the **Figure 3** we compare the paraelectric and ferroelectric soft phonon spectra of PbTiO$_3$ calculated by us with experimentally observed by Shirane et al [41]**.** Parameters corresponding to the best fitting of PbTiO$_3$ spectra are listed in the last row of the **Table I.** It is clear from the figure that only the solid curves calculated for both nonzero static and dynamic flexocoupling constants ($f_{11}= -8$ V, $f_{44}= -1.9$ V and $M_{11}= -2\times10^{-8}$ V s$^2$/m$^2$) describe quantitatively observed paraelectric and ferroelectric soft phonon spectra of PbTiO$_3$ at small **k** (compare dotted, dashed and solid curves in the **Figures 3**). Therefore it is hardly possibly to fit the experimental results within a proposed analytical approach but without inclusion of nonzero static and dynamic flexocoupling constants. Hence we conclude that both the static and dynamic contributions are critically important to describe quantitatively the available experimental data.

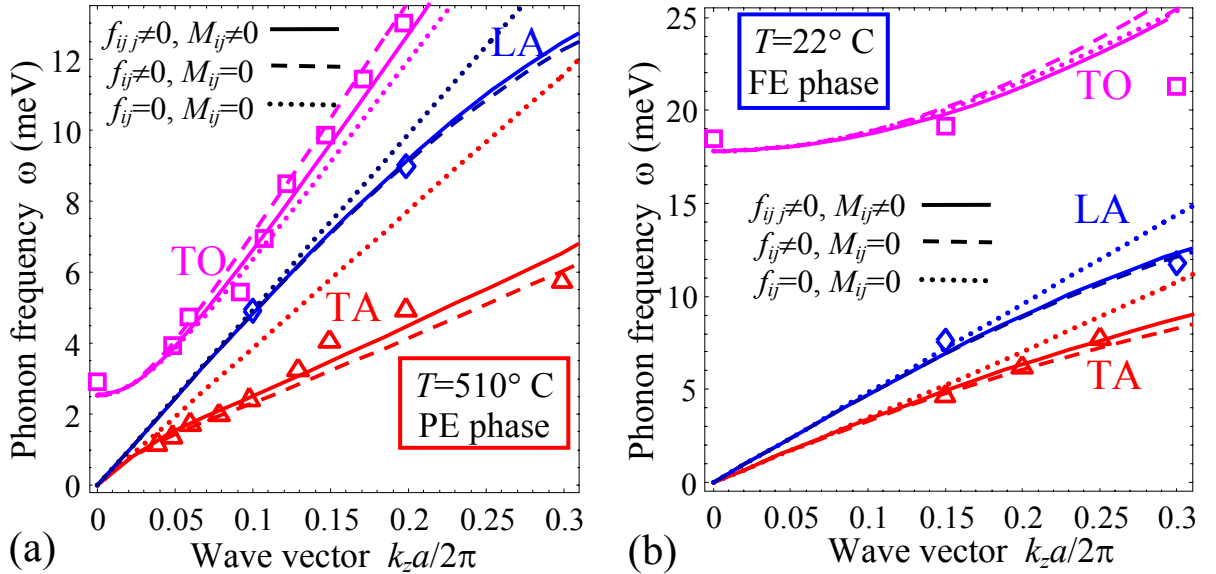

**Figure 3**. Soft phonon branches calculated in PbTiO$_3$ vs. $k$ in units $\pi/a$, $a$ is a lattice constant. Plot **(a)** corresponds to the paraelectric (PE) phase of ($T$=510 C), plot **(b)** is calculated in the ferroelectric (FE) phase ($T$=22 C) for the case $\vec{k} \uparrow\uparrow \vec{P}_S$. Symbols are experimental data from ref.[41]. Dotted curves are calculated without flexoelectric coupling ($f_{ij}$=0 and $M_{ij}$=0); dashed curves are calculated with the static but without the dynamic effect ($f_{ij}\neq 0$ and $M_{ij}$=0) and solid curves are calculated with the dynamic and static flexoelectric effects included ($f_{ij}\neq 0$ and $M_{ij}\neq 0$). Parameters corresponding to the best fitting of PbTiO$_3$ spectra are listed in the last row of the **Table I.**

## VI. Summary

Within Landau-Ginzburg-Devonshire approach we establish the impact of the flexocoupling on the correlation function of the long-range order parameter fluctuations in ferroelectric phase of ferroics with local disordering sources and obtained analytical expressions for the generalized susceptibility and phonon dispersion relations for ferroelectrics with arbitrary symmetry, elastic and electrostrictive anisotropy. Relatively simple analytical expressions for the susceptibility components and soft phonons dispersion law are available in the cubic approximation for the elastic properties of ferroelectric. We studied their physical manifestations, namely:

a) The joint action of static and dynamic flexoelectric effect induces non-diagonal components of generalized susceptibility, which amplitude is proportional to the convolution of the spontaneous polarization with flexocoupling constants.

b) The flexocoupling essentially broadens the k-spectrum of generalized susceptibility and so decreases the correlation radii

c) The contribution of spontaneous polarization via ferroelectric nonlinearity and electrostriction mechanisms can lead to both broadening and narrowing on the susceptibility k-spectrum.

d) The spontaneous polarization appearance leads to the removal of the modes degeneration and consequently to the appearance of different transverse acoustic and optics modes. The flexoelectric coupling significantly increases the splitting of the acoustic modes, as well as leads to the additional "pushing away" of the optical and acoustic soft mode phonon branches.

e) It appeared hardly possible to fit adequately the experimentally observed phonon spectra at small k within a proposed analytical approach for zero static and dynamic flexocoupling constants. Hence we conclude that both the static and dynamic contributions are critically important to describe quantitatively the available experimental data.

Finally, we would like to underline that we did not aim to reach the quantitative agreement between the calculated and experimentally observed soft phonon spectra entire the first Brillouin zone. Consideration of the problem for higher **k** values requires including the anharmonicity and higher gradient terms to modify the used harmonic approach. However our results prove the evident importance of the static and dynamic flexocouplings for the adequate description of the generalized susceptibilities and soft phonon spectra near the centre of the Brillouin zone. Since modern and classic experimental methods readily capture the small-k region further study of the flexocouplings impact on the susceptibility spectra for all crystallographic symmetries seems important. These results can be principally important for quantitative analyses of the experimental data broad spectrum including neutron and Brillouin scattering, which collects unique information from the structural factors and phonon dispersion.

**Acknowledgements.** Authors gratefully acknowledge extremely useful suggestion to include the dynamic flexoelectric into the theoretical consideration, multiple consultations and discussions with Prof. A.K. Tagantsev (EPFL). E.A.E. and A.N.M. acknowledge National Academy of Sciences of Ukraine (grant 35-02-15).